\documentstyle[prl,preprint,aps]{revtex}

\begin{document}
 \tolerance 50000

\draft

\title{Phase Diagram of the  2-Leg Heisenberg Ladder \\
with Alternating Dimerization} 
\author{  
M.A. Mart\'{\i}n-Delgado$^{1}$,
J. Dukelsky$^{2}$ and 
G. Sierra$^{3}$
 } 
\address{ 
$^{1}$Departamento de
F\'{\i}sica Te\'orica I, Universidad Complutense. Madrid, Spain.
\\ 
$^{2}$Instituto de Estructura de la Materia, C.S.I.C.,Madrid, Spain.
 \\
$^{3}$Instituto de Matem\'aticas y F\'{\i}sica Fundamental, C.S.I.C.,
Madrid, Spain. 
}

\maketitle 
\widetext

\vspace*{-1.0truecm}

\begin{abstract} 
\begin{center}
 \parbox{14cm}{Using the Lanczos method 
we determine the phase diagram of the 2-leg
AF-Heisenberg ladder with alternating dimerization.
It consists of a resonating valence bond phase
and a dimer phase separated by a critical line.
Our results are in good agreement with previous conjectures
obtained with the non linear sigma model. }
\end{center}
\end{abstract}

\pacs{
 \hspace{2.5cm} 
}

\narrowtext

In the last few years,
Heisenberg spin ladders have attracted considerable 
attention \cite{dagotto-rice} due
to possible connections to materials 
exhibiting high T$_c$ superconductivity.
Theoretical studies \cite{varios} 
hint at the possibility that even-legged ladders
become superconductors when doped with charge carriers.  There is
some experimental support for 
this possibility as Sr$_{0.4}$Ca$_{13.6}$Cu$_{24}$O$_{41.84}$,
a material with spin-${1 \over 2}$ chains and 2-chain ladders, was 
shown to superconduct at 12 K and 3 GPa  \cite{uehara}.

Spin ladders are also interesting theoretically because of their
unexpected behavior when viewed as interpolating constructions between the 
spin-${1 \over 2}$ 1D antiferromagnetic Heisenberg chain and their
2D square analog.  The latter is ordered 
at low temperatures \cite{manu},
while the former  exhibits spin-spin correlations
 with a power-law decay along the 1D chain.  
Yet the transition between
these limits by forming spin-${1 \over 2}$ ladders with increasing
numbers of legs is not smooth: ladders with even numbers of legs 
have a finite gap to the lowest triplet state and an exponential
decay of spin-spin correlations, while odd-leg ladders have gapless 
excitations and a power-law fall off of spin-spin correlations.
These issues and examples of materials exhibiting these properties
are discussed in several recent reviews  \cite{dagotto-rice}(and references therein).

With the hope of gaining a deeper insight into such systems, 
a variety of techniques have 
been employed, such as
exact diagonalizations with the Lanczos 
algorithm \cite{barnes}, quantum Monte
Carlo simulations \cite{barnes}\cite{greven}, 
and density matrix renormalization group 
methods  \cite{White}\cite{varios}.   
Lanczos calculations, while limited to small size $L$ of the lattice, 
play an important role in testing 
approximate schemes \cite{nos2} and also in evaluating dynamic quantities that are difficult to treat in other
approaches.

In this letter we present a Lanczos \cite{cw} \cite{review}
study of the 2-leg AF-Heisenberg ladder
with alternating dimerization. By alternating dimerization we mean that
the exchange coupling constants along the legs are not uniform in strength,
but change according to a definite pattern shown in Fig.1a.
The choice of this pattern among many other possibilities is not 
arbitrary, as was shown in reference \cite{nos} where a great variety of 
quantum critical phases were conjectured 
and classified using the Haldane mapping of 
spin systems into the non linear sigma model. 
The columnar dimerization of the two legged ladder
has been considered in reference \cite{tot},
where it was shown that a critical line emerges 
for ferromagnetic inter-leg couplings and 
antiferromagnetic intra-leg ones.

Here we present Lanczos
results for the singlet-triplet gap in
$2 \times L$ systems with  $L$ = 4, 6, 8, 10 and 12.
This in turn allow us to estimate the critical curve in the 
space of couplings of the dimerized ladder which, as 
will be shown below, is two dimensional.
We also compare the numerical results with the 
analytical predictions based on perturbation theory 
and on the non linear sigma model\cite{nos}.

To be more precise, let us start with the 
 Hamiltonian for the spin-${1 \over 2}$ Heisenberg spin ladder
consisting of two coupled chains with alternating couplings
given by,

\begin{equation}
H = \sum_{a=1}^2 \sum_{n=1}^{L} J \;\; (1 + (-1)^{a+1}\gamma)
\; \; \vec{S}_a(n) \cdot \vec {S}_a(n+1) + 
J^\prime \sum_{n=1}^L
\vec{S}_1(n) \cdot \vec{S}_2(n),
\label{1}
\end{equation}

\noindent where $\vec{S}_a(n)$ denotes the quantum spin-$1/2$ at site $n$ in the 
leg $a=1,2$ of the ladder, $J$ and $J^\prime$ are coupling constants 
along the legs and the rungs respectively, and $\gamma$
is the dimerization parameter.
In order to keep the system always in the antiferromagnetic regime,
we restrict the range of variation as $|\gamma| \leq 1$ (see figure 1a)).
We use periodic boundary conditions along the legs of the ladder.

In the strong coupling limit (  $J^\prime/J \gg 1$)
and in the absence of staggering ($\gamma=0$),
the spins across each rung form a singlet and the 
ground state wave function is the coherent superposition
of these singlets
\cite{barnes}.  The
ground state thus has $S$ = 0 and an energy gap proportional to
$J^\prime$, with perturbative corrections of relative size
$J/J^\prime$. These perturbative expansions can be extended to the
case of non-vanishing dimerization $\gamma \neq 0$.
From second order perturbation theory we obtain for the gap,

\begin{equation}
\Delta (\gamma,J^\prime/J) = J^\prime -J + 
({1\over 2} - 2\gamma^2)\; {J^2\over J^\prime}  
\label{2}
\end{equation}

\noindent Playing with the two coupling constants present in
the problem, we can tune for instance the dimerization in order to have 
a vanishing spin gap $\Delta (\gamma_c,(J^\prime/J)_c) = 0$.
This yields the existence of a critical curve for gapless excitations
which in this approximation reads as follows,

\begin{equation}
(J^\prime/J)_c = {1\over 2} (1 + \sqrt{8\gamma_c^2 - 1}),
\label{3}
\end{equation}

A more elegant and powerful method was used in \cite{nos} to determine
the existence of critical curves for any spin $S$ and number of legs $n_l$
in the two-dimensional plane of coupling constants $\gamma$ and $J^\prime/J$.
There we employed the well-known Haldane mapping \cite{Hal}
 of a AF-Heisenberg 
model to the NLSM (Non-Linear Sigma Model). In this approximation,
the existence of gapless excitations is detected through the condition
that the $\theta$-parameter associated to the topological term of the NLSM
is an odd multiple of $\pi$. In the case of a $n_l=2$ ladder with spin-1/2
degrees of freedom we found,

\begin{equation}
\theta = 2 \pi (1 - {\gamma \over 1 + {J^\prime/2J}}),
\label{4}
\end{equation}

\noindent The criticality condition $\theta=\pi$ leads in this framework
to the existense of a critical curve given by,

\begin{equation}
\gamma_c = \frac{1}{2} (1 + {J_c^\prime/2J_c} ),
\label{5}
\end{equation}

\noindent which is a straight line starting from $J^\prime/J=2$
at $\gamma=1$ and ending at $J^\prime/J=0$ and $\gamma = 1/2$.
The NLSM has to be taken as a guide to the physics of the problem
but in general it is not quantitatively correct. 
Indeed at $J'/J =0$ the critical curve has to end up at
the point $\gamma =0$, corresponding to two uncoupled uniform
chains. Thus the point $(J'/J=0, \gamma=1/2)$ is certainly 
non critical.
On the other hand,   
the point $(\gamma=1,J^\prime/J=2)$ corresponds to the exact
solution
as shown in reference \cite{nos}.  
The reason for this is essentially geometrical as shown
in figure 1c. At $\gamma=1$ the 2-leg ladder becomes 
effectively a single spin chain with the shape
of a snake. If moreover $J'/J$ equals 2 this chain 
becomes uniform
and consequently is gapless. Away from this ratio
there appears a standard dimerization gap \cite{cross-fisher}.

These strong-coupling arguments together with the
weak coupling arguments given in reference \cite{nos}
suggest the existence of
a critical curve  connecting the points 
$(J'/J,\gamma) = (2,1)$ and $(0,0)$.

However, despite all this mounting evidence in favor of the existence of
a critical curve in the two-dimensional space of coupling constants,
it is apparent that a direct determination of this phase diagram
must be reported.

Here we carry out a Lanczos diagonalization on finite $2\times L$--site
lattices in order to compute the spin gap
$\Delta_L(\gamma, J'/J)$ for a range of values of the coupling constants
given by $0 \leq J'/J \leq 2$ and $0 \leq \gamma \leq 1$. 
 
In Fig.2 we show a 3D-plot of the surface of gaps 
$\Delta(\gamma,J^\prime/J)_L$ for $L=10$ where we can inequivocally 
appreciate a steep valley signaling the existence of the critical 
curve we are searching for.

Next we perform a finite size analysis to determine
the critical and non-critical regions in the space of coupling 
constants.

In a gapped regime
the spin gap  has the 
scaling behaviour  \cite{barnes}

\begin{equation}
\Delta_L - \Delta_\infty = C {e^{-L/\xi} \over L}
\label{6}
\end{equation}

\noindent
where $\xi$ is the correlation length.  
If $\xi$ is larger than the corresponding lattice size 
one can actually get a better fit of  the data with the formula

\begin{equation}
\Delta_L - \Delta_\infty = {C  \over L}
\label{7}
\end{equation}

\noindent where $\Delta_\infty$ should be a small
quantity consistent with the large value of $\xi$. 

For a given value of $J'/J$ between 0 and 2 we have
varied $\gamma$ from 0 to 1. In this interval of
$\gamma$ we find regions with a massive behaviour of the
gap described by (\ref{6}) with $\xi$ smaller than the lattice
size, together with small windows where the scaling is
better described by (\ref{7}) with $\Delta_{\infty}$ 
varying between 0.04 and 0.07. 
The critical point is associated with the minimum 
in this region. The latter 
points are depicted in figure 3 and  we associate
them with the critical line conjectured in ref \cite{nos}. 
Strictely speaking, using the Lanczos method with
$L$ up to 12, we cannot
rule out a very small gap in  the region we believe is critical.
However we think  
that this possibility is remote ( see ref \cite{hida} for
similar difficulties in detecting gapless phases for the
two leg ladders with ferromagnetic interchain couplings).

In figure 3 we have plotted our numerical results together
with the continuous curve given by $ J^\prime/J=2 \gamma^{2/3}$ which
gives a reasonable overall fit. 

We can motivate this curve as follows.
A dimerized spin 1/2  chain has a gap  
which behaves as $ J \gamma^{2/3}$
according to the  Cross and Fisher law.
This law can be proved using Conformal 
Field Theory. Indeed a  
dimerized spin 1/2 chain is described 
by a  $SU(2)$ WZW model 
perturbed 
by a relevant operator
with conformal dimension 1/2, 
which by the way is the unique relevant
operator available in the model\cite{Affleck}. 
On the other hand a uniform spin 1/2 ladder, in the
weak coupling limit $J/J' <<1$,  is described by 
two WZW models, corresponding to every chain,
which are coupled through a relevant
operator with dimension 1 \cite{coupled}. This leads
to a linear dependence on $J'$ of  the spin gap of uniform chains. 
The critical curve that we propose, i.e. $J'/J= 2 \gamma^{2/3}$,
is an interesting combination between  these two 
laws, which could be called the Cross-Fisher-ladder (CFL) law. 
From the
previous conformal field theory arguments we see that
the mass scale induced by both $J'$ and $\gamma$ couplings behave
as $mass \sim J \; \gamma^{2/3} \;\sim \; J'$ and since
the model is critical these two quantities should be identical.  
The proportionaly factor 2 in the $CFL$ law is fixed by the
fact that at $\gamma=1$ the dimerized ladder becomes
a chain with the shape of a "snake'' (see figure 1b)
\cite{nos} , which is critical for 
$J'/J= 2$.  
Thus the two mechanisms for generating a spin gap cancel each other.

The critical curve  of figure 3 separates two phases which can be neatly
identified with the resonating valence bond phase of ladders (RVB) 
and the dimer phase of spin chains. We can try to understand why
there should be a critical line between these two different regimes.
On the RVB region the basic mechanism which lowers the energy
is the resonance between parallel bonds on both rungs and legs
which is proportional to $J'$. 
In the dimer phase the bonds are fixed to staggered
configurations and their energy is proportional to $J \gamma$.
When both $J'$ and $\gamma$ are non vanishing there is
a competition between 
these two types of configurations: resonance is favored by $J'$
but disfavored by $\gamma$, while staggering is favored by $\gamma$
but disfavored 
by $J'$. The critical behaviour  appears when  there is a perfect
balance between the two phenomena.

It may seem that the RVB and dimer phases of the alternating
ladder are radically different.
However, as shown in \cite{hida,white2},
one can connect the dimer and the RVB phases
with the Haldane phase of a spin 1 chain. This is not inconsistent
with our results. All that says is that the path that connects
the RVB and dimer phases of the dimerized 2-leg ladder 
lies outside our two dimensional 
phase diagram.

In summary, we have addressed 
the problem of whether the dimerized ground state
survives the quantum interchain and alternating
staggering fluctuations.
We have found that the Lanczos's numerical results  confirm the
existence of a critical phase where the 2-leg ladder remains gapless.
This justifies that the ``snake 
mechanism" presented before is also valid away of the 
strong-coupling/strong-staggering 
regime (the upper-right part of the phase diagram.)


                Note added: After completion of this paper
we have been aware of the paper of Flocke \cite{Flocke} which
agrees essentially with our results. However a theoretical
explanation of the phase diagram as well as the law $\gamma^{2/3}$
for the critical curve separating the dimer and the RVB phases
are lacking in \cite{Flocke}.


{\bf Acknowledgements}: We would like to thank R. Shankar
for discussions in the early stages of this work. 
We are also grateful to H. Nishimori for the
package  TITPACK version 2 which we have used 
for the diagonalization of spin-1/2 systems.

MAMD and GS acknowledges support from the 
DIGICYT under contract No. PB96/0906.
JD acknowledges support from the DIGICYT under
contract No. PB95/0123

\newpage

\section*{Figure Captions}

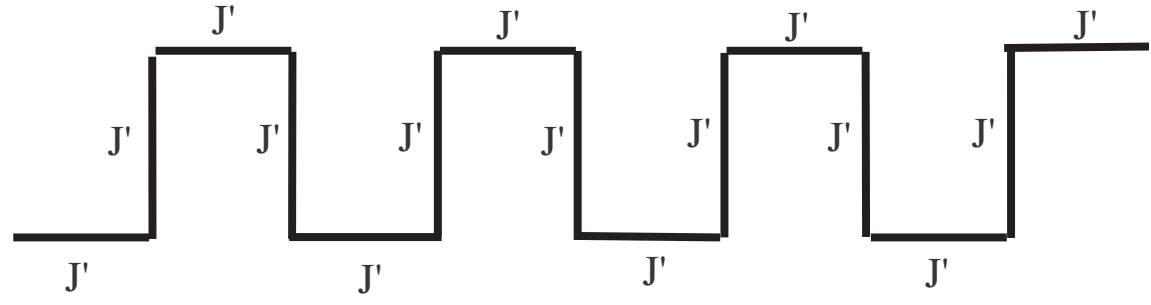
{\bf Fig.1} a) A 2-leg ladder with alternating dimerized couplings
$J(1\pm \gamma)$ along the horizontal legs and $J^\prime$ along
the vertical rungs.
b) A maximally dimerized ladder with $\gamma=1$.
c) A uniform 1D-Heisenberg chain emerging from the ``snake" pattern when
$\gamma=1$ and $J^\prime/J=2$.

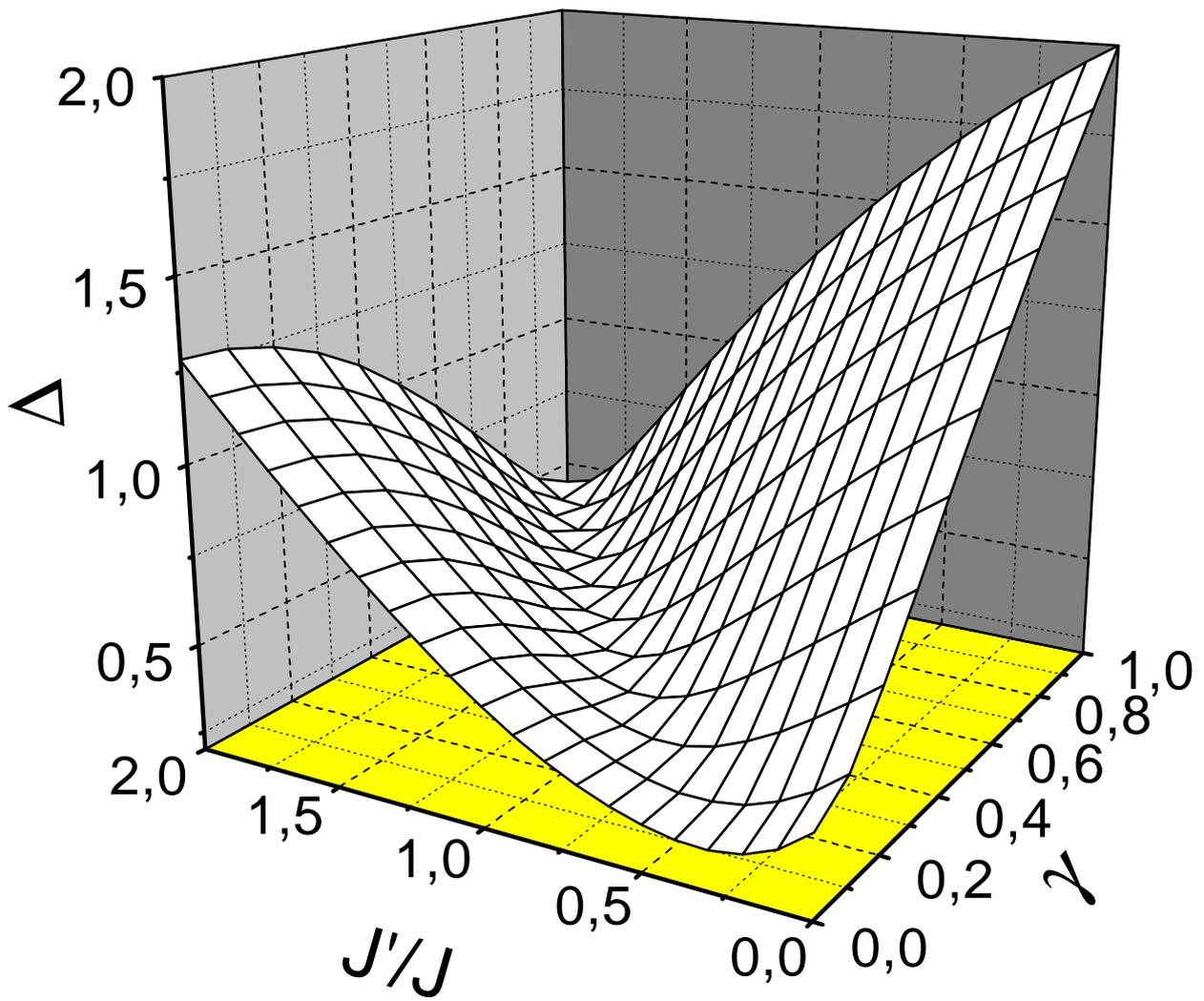
{\bf Fig.2} A 3D-plot of the surface of gaps 
$\Delta(\gamma,J^\prime/J)_L$, for $L=10$, in the two-dimensional
parameter space of couplings showing the existence of a steep valley
which we associate to the critical curve in the phase diagram.

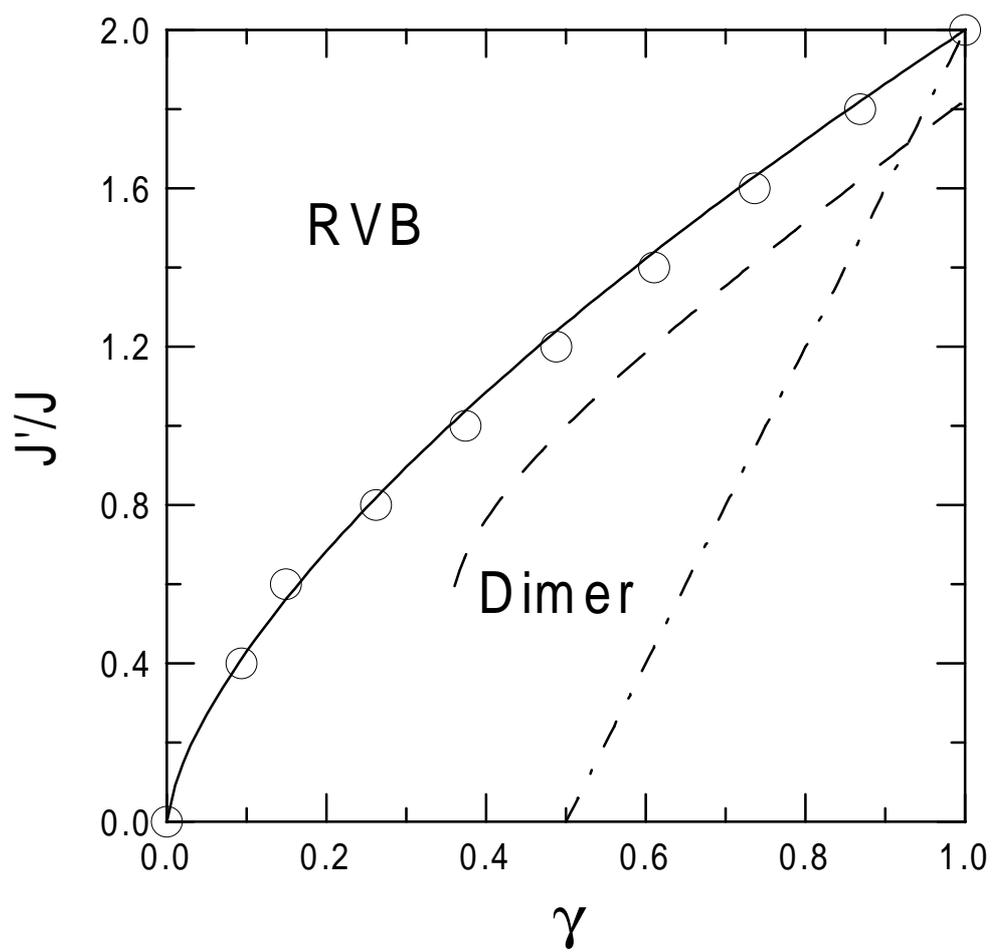
{\bf Fig.3} Phase diagram resulting from the analysis of the Lanczos
results. The circles denote the extrapolated numerical results.
The solid curve is the function $2\gamma^{2/3}$ (explained in the text)
which fits those circles
quite well. RVB and Dimer phases are separated by this line.
The dashed line corresponds to the 
perturbation result given in eq.(\ref{3}), while the dotted-dashed
line corresponds to the NLSM result given in eq.(\ref{5}).

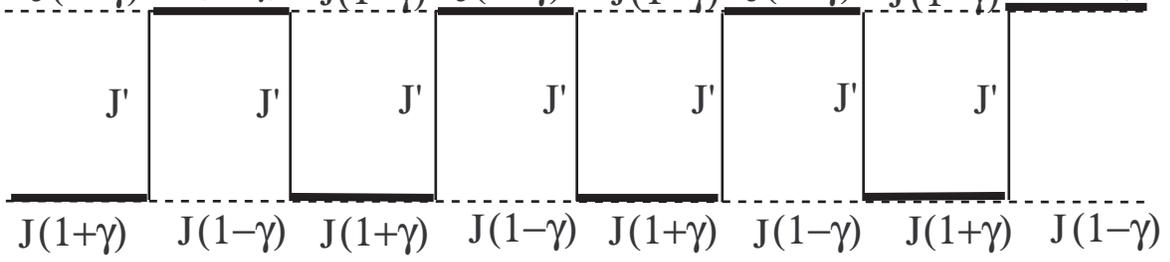
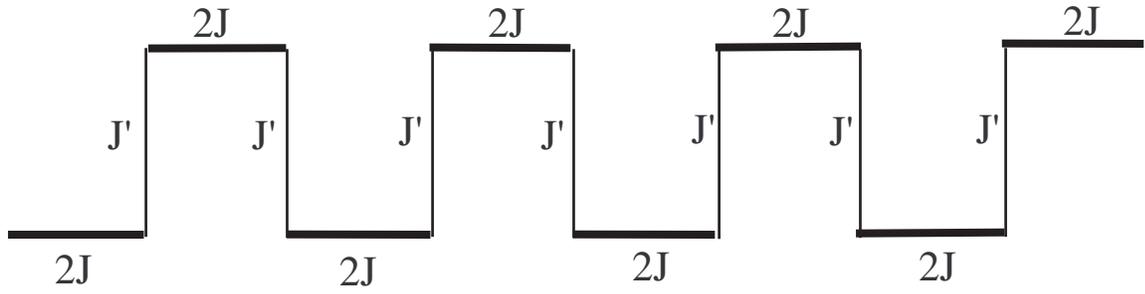
\end{document}